\documentclass{PoS}

\PoS{PoS(LAT2005)219}
\usepackage[dvips]{graphicx}

\title{B meson decays at high velocity from mNRQCD }

\ShortTitle{B meson decays at high velocity from mNRQCD }

\author{\speaker{A. Dougall}$^{a}$, Kerryann M. Foley$^b$,  
  C. T. H. Davies$^{a}$ and G. Peter Lepage$^{b}$\\ \\
  \llap{$^a$}Department of Physics and Astronomy, University of
  Glasgow, Glasgow G12 8QQ, UK\\
  E-mail: \email{adougall@physics.gla.ac.uk} \\
  \llap{$^b$}Newman Laboratory for Elementary-Particle Physics,
  Cornell University, Ithica, NY 14853, USA}

\abstract{The moving NRQCD (mNRQCD) formalism facilitates the
  simulation of heavy meson decays at large recoil. We present
  preliminary results for a number of quantities including the energy
  splittings and the $B$ meson decay constant that demonstrate
  that the formalism is accurate over a range of momenta.}

\FullConference{XXIIIrd International Symposium on Lattice Field Theory\\
  25-30 July 2005\\
  Trinity College, Dublin, Ireland}

\begin{document}

\section{Introduction}

At present there is much work being done to establish accurate values
for certain CKM matrix elements and this requires the comparison of
theory and experiment for exclusive semileptonic decays of heavy
mesons. For example, a value for the CKM matrix element $V_{ub}$  can
be obtained from the exclusive semileptonic heavy decay  $B\rightarrow
\pi l \nu$ but  only when it is combined with the corresponding QCD
form factor. Theoretically, the calculation of such form factors
necessitates a non-perturbative handling of QCD, to which the lattice
is ideally suited. Indeed, current lattice simulations work well
within the low recoil region (high $q^2$). However, gaining access to
the full kinematic regime is problematic due to the fact that the
final state meson can have large recoil momentum in the rest frame of
the initial heavy meson, resulting in large discretisation errors and
noisier signals that increase the statistical error. Thus comparisons
between theory and experiment have been restricted to the low recoil
region. Accessing the form factors at high recoil is particularly
important since much of the experimental data lies within this
kinematic region. 

The ultimate goal of this work is therefore to extend the form factor
calculations to cover the entire kinematic range. These techniques can
be applied to different semileptonic decays, but here we focus on the
most extreme case, $B \rightarrow \pi l \nu$. Lattice calculations are
usually carried out in the rest frame of the $B$ meson, however, the
approach in mNRQCD is to treat the $B$ meson in a moving frame, whilst
maintaining control over the systematic errors. This is achieved by
finding the optimal reference frame in which the discretisation and
statistical errors are minimised, and then using a mNRQCD action to
prevent large discretisation errors associated with the $B$ meson. The
central argument for this approach is that although the heavy quark
carries almost all the heavy meson momentum, the internal dynamics of
the meson are essentially non-relativistic. Previous work in this field
includes \cite{Hashimoto:1995in,Mandula:1997hb,Sloan:1997fc}.

\subsection{Reference frame}

For the most extreme kinematic case, $q^2(=p_B^2 - p_\pi^2)=0$,  we
need to find a frame that minimises both the statistical errors that
depend on the average kinetic energy, and the discretisation errors
that depend on the momenta. In the case of the statistical errors, the
frame in which  the kinetic energy is at a minimum is ${\bf p_B} =
{\bf p_\pi}$.  However, it is possible to have ${\bf p_\pi}$ lower as
larger values for ${\bf p_B}$ can be dealt with using more elaborate
sources to reduce the statistical error. The discretisation errors are
analysed by considering the momentum flow within the $B$ and $\pi$
mesons and the optimal frame lies within these bounds.

By choosing a reference frame in which the $B$ meson is moving, this
reduces the error associated with the pion's momentum, but increases
the error from the $B$ meson's momentum. However, most of the $B$
meson momentum is carried by the $b$ quark and so we can construct a
moving NRQCD Lagrangian to deal with this.  To minimise errors
associated with the light quark action and light meson momentum, it is
essential that the light quark action is $O(a^2)$ improved.

\subsection{mNRQCD}

In constructing the {mNRQCD} Lagrangian, the total momentum of the $b$
quark is defined as $P_b^{\mu}={m_b u^{\mu}} + {k^{\mu}}$, where the
first term is the {\it average} momentum and $u^\mu = P_B/m_b = \gamma
(1,{\bf v})$ is the  4-velocity. The second term, $k^{\mu}$ is the
{\it residual} momentum, which is of the same order as the light
quark's momentum. The strategy is to treat the $b$ quark's
average momentum exactly, whilst discretising the residual
momentum. The mNRQCD Lagrangian can be derived in two ways. One can
either start with the NRQCD Lagrangian (derived from the FWT
transformation on the usual relativistic quark fields) and boost the
fields, or, one can re-write the Lagrangian in terms of boosted fields
and then perform the FWT transformation, we outline the
latter. Further details are given in Ref. \cite{paper}.

In the NRQCD formalism, the FWT transformation is applied to the
quark fields in order to decouple the quark from the anti-quark,
greatly simplifying the numerics. It also removes the rest energy from
the Lagrangian, thus leading to smaller discretisation errors. The
decoupling is performed as an expansion in the heavy quark mass or the
internal quark velocity and is optimal for small 3-momentum. However,
the heavy quark in $B$ decays has very large 3-momentum that 
is boosted with respect to the lattice frame ($P_b$) and therefore it 
is not possible to apply a FWT transformation on the fields. However,
by re-writing the Lagrangian in terms of boosted fields and operators,
a FWT transformation can then be applied\\

\hspace{2cm}
$\Psi = \frac{\Lambda({\bf v})}{\sqrt{\gamma}}
Te^{-imu.x}A_{D_t}
\left(
  \begin{array}{c}
    \psi_v\\
    \chi_v
  \end{array}
\right)
$; 
\hspace{1cm}$
\Lambda = \frac{1}{\sqrt{2(1+\gamma)}}
\left( 
  \begin{array}{cc} 
    1+ \gamma& \sigma.{\bf v}\\
    \sigma.{\bf v}& 1+\gamma
  \end{array}
\right)
$ \\

\noindent
to give the mNRQCD Lagrangian\\
\begin{eqnarray*}
  L = \psi^\dagger \left(
    iD_t + i{\bf v.D} + \frac{{\bf D}^2}{2\gamma m}
    - \frac{({\bf v.D})^2}{2\gamma m} +
    \frac{{\bf \sigma.\tilde{B}}}{2 \gamma m} 
    \ldots \right) \psi.
\end{eqnarray*}

\vspace{0.5cm}
\noindent
In addition to the usual terms appearing in the NRQCD Lagrangian,
there are now terms that are a function of the 3-velocity of
the $b$ quark, ${\bf v}$.

\section{Details of current work}

Here we define the time evolution operators that are currently being
tested and outline the details of the lattice calculation.

\subsection{mNRQCD action}

\noindent
From the mNRQCD Lagrangian, the time evolution operator is given by\\ 
\begin{eqnarray*}
  {\bf H} = \left(1 - \frac{
      {aH_0}}{2n} \right)^{n}_{t}
  \left( 1 - \frac{
      {\delta H}}{2} \right)_t
  U_{4, t-1}^\dagger
  \left( 1 - \frac{
      {\delta H}}{2} \right)_{t-1}
  \left(1 - \frac{
      {H_0}}{2n}\right)^n_{t-1}
\end{eqnarray*}

\noindent    
Early tests have been carried out for the simplest action, where
\begin{eqnarray*}
  {H_0} 
  &=& -\frac{\Delta^{(2)}}{2m_Q} -
  {\bf v.} \Delta^{(\pm)} + 
  ({\bf v.} \Delta^{(\pm)})^2 \left(
    \frac{c_v}{2m\gamma} + \frac{c_{ct}}{2n} \right),
\end{eqnarray*}
\noindent
and tests are now being carried out on the fully $O(1/m)$ improved case, where
\begin{eqnarray*}
  {\delta H} &=& 
  \frac{{\bf \sigma.\tilde{B}}}{2m\gamma}  - 
  \frac{c_{c,1}}{m\gamma} \{ {\bf v.}\Delta^{(\pm)}, \Delta^{(2)} \} - 
  \frac{c_{c,2}}{m\gamma} \left({\bf v.}\Delta^{(\pm)}\right)^3 + 
  \frac{c_{c,6}}{6} v_i \Delta_i^{+} \Delta_i^{\pm} \Delta_i^{-} 
  \\ \nonumber &&\,\, 
  \frac{c_{c,7}}{m\gamma} \Delta^{(\pm)^{(4)}} +
  \left(
    \frac{c_{c,8}}{m\gamma} + \frac{c_{c,9}}{n_1} \right)
  \{ {\bf v.}\Delta^{(\pm)}, v_i \Delta_i^+ \Delta_i^{\pm}
  \Delta_i^- \}.
\end{eqnarray*}

\noindent
Note that all the coefficients (labelled $c$) are 1 at tree-level.

\subsection{Simulation details} 

This work was carried out on a $12^3 \times 24$ lattice with inverse
lattice spacing $a^{-1}=1.36(14)$GeV (set from Upsilon splitting) for
a data set of 243 quenched configurations. The light quark propagators
were generated using clover fermions ($\kappa \sim
\kappa_s$). Gaussian smearing was used at the source and both local
and Gaussian smearing was used at the sink. The heavy quark
propagators were generated using both the simplest and $O(1/m)$
improved action. Local and APE smearing was used at both source and
sink. The heavy quark mass ($am_Q$) ranges from 2-8, where 4
corresponds to the mass of the $b$ quark on this lattice and we looked
at a range of velocities, ${\bf v}/c=0-0.8$.

\section{Perturbation theory}

As with NRQCD, mNRQCD is constructed from a set of non-renormalisable
interactions, each of which is accompanied by a coefficient which must
be calculated perturbatively in order to obtain physical results. The
one-loop parameters which make up the self energy include the energy
shift, mass and wavefunction renormalisation and also, the velocity
renormalisation which results from the division of the total heavy quark
momentum into average and residual pieces. A remnant of
reparameterisation invariance on the lattice prevents this coefficient
from becoming too large. The inverse quark propagator can be written as   
a combination of the free propagator and the heavy quark's self
energy,
\begin{eqnarray*}
  G^{-1} &=& Q^{-1} - a\Sigma(k)\\
  &=& -ik_4 - \alpha_s\Omega_0+\alpha_s ik_4 \Omega_1+{\bf v.k}
  - \alpha_s {\bf v.k} \Omega_v +\ldots\\
  &=& {Z_\psi}
  \left(-i\bar{k_4} + \frac{{\bf k}^2}{2\gamma_R m_R}
    + \frac{{\bf P_R .k}}{\gamma_R m_R} 
    +\ldots \right)
\end{eqnarray*}

\noindent      
where the renormalisation constants, eg. ${Z_\psi}$, are expressed in
terms of the coefficients, $\Omega_n$, appearing in the self
energy. The value of each $\Omega_n$ is obtained by differentiation,
further details of which are given in Ref. \cite{Dougall:2004hw}. 

\section{Testing mNRQCD}
      
\begin{figure}
  \begin{minipage}[l]{0.48\linewidth}
    \vspace{-0.5cm}
    \begin{center}
      \includegraphics[width=0.95\linewidth]{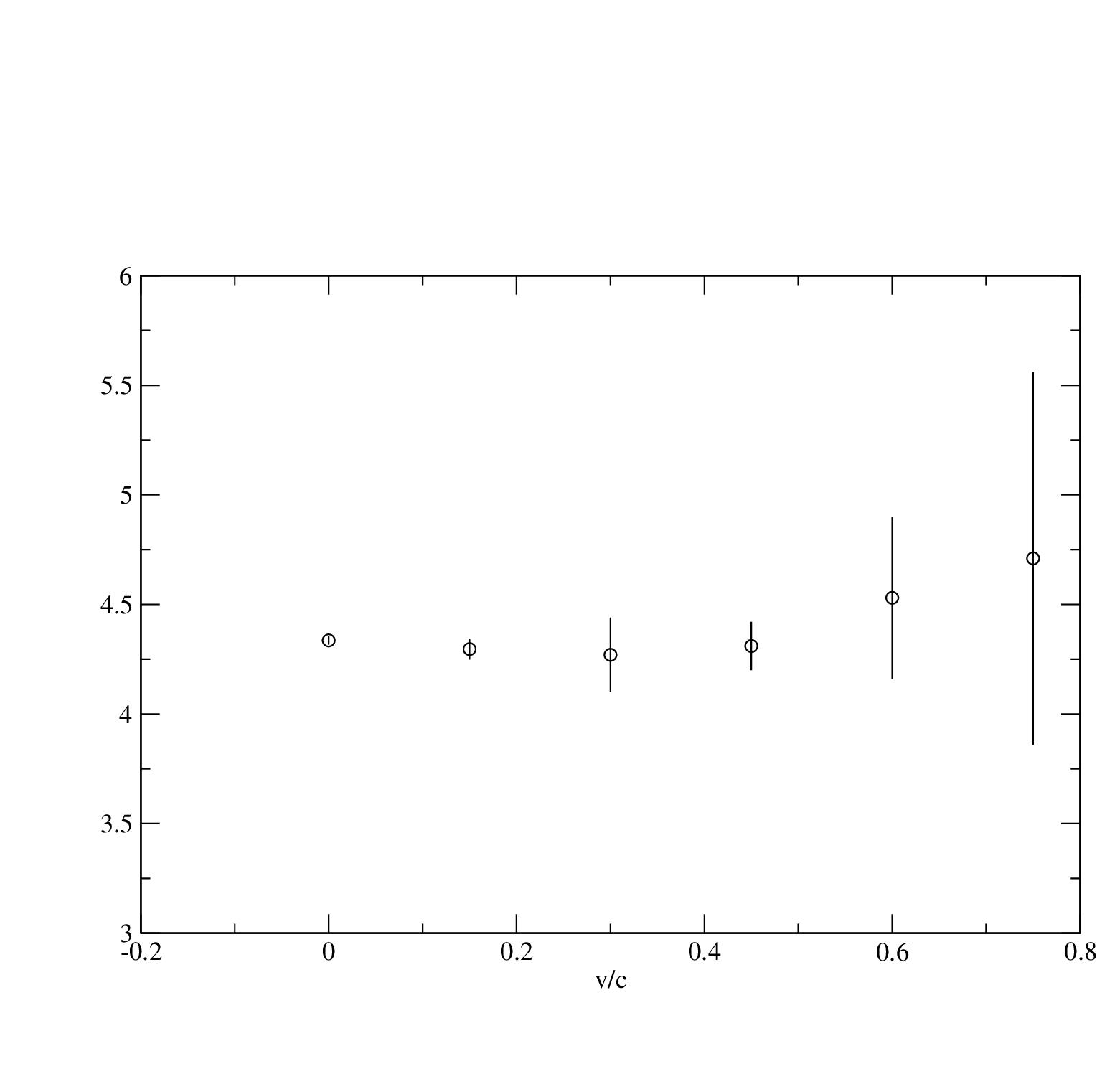}
      \caption{The HH kinetic mass as a function of the
        velocity from the simplest mNRQCD action.} 
      \label{mkin2}
    \end{center}    
    
  \end{minipage}
  \hspace{3mm}
  \begin{minipage}[r]{0.48\linewidth}
    \begin{center}
      \includegraphics[width=0.95\linewidth]{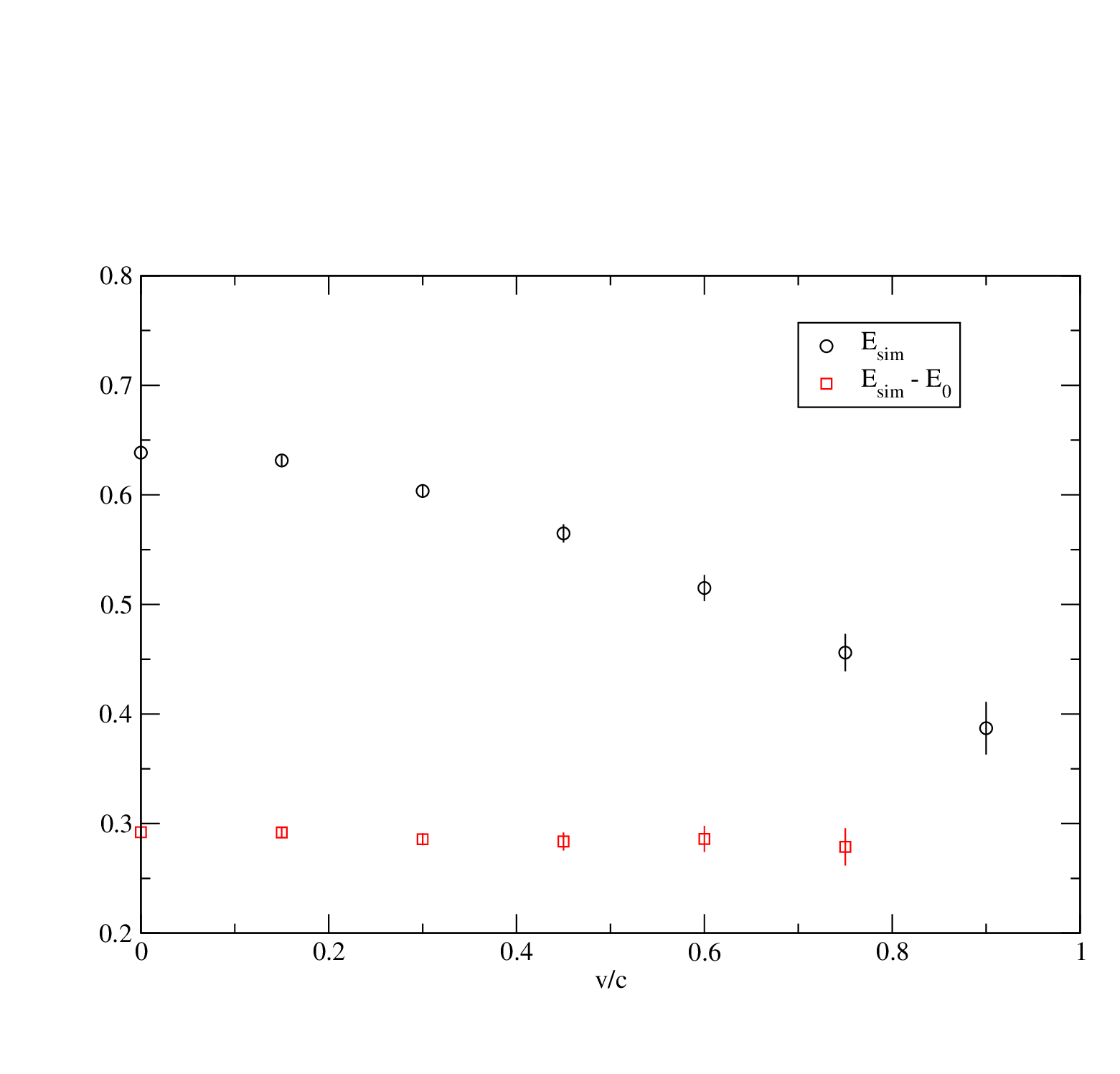}
      \caption{The HL ground state and binding energy at zero
        residual momentum. Note that binding energy is constant with
        increasing velocity.}
      \label{Esim}
    \end{center}    
    
  \end{minipage}
\end{figure}

\begin{figure}
  \begin{minipage}[l]{0.48\linewidth}
    \vspace{-2cm}
    \begin{center}
      \includegraphics[width=0.95\linewidth]{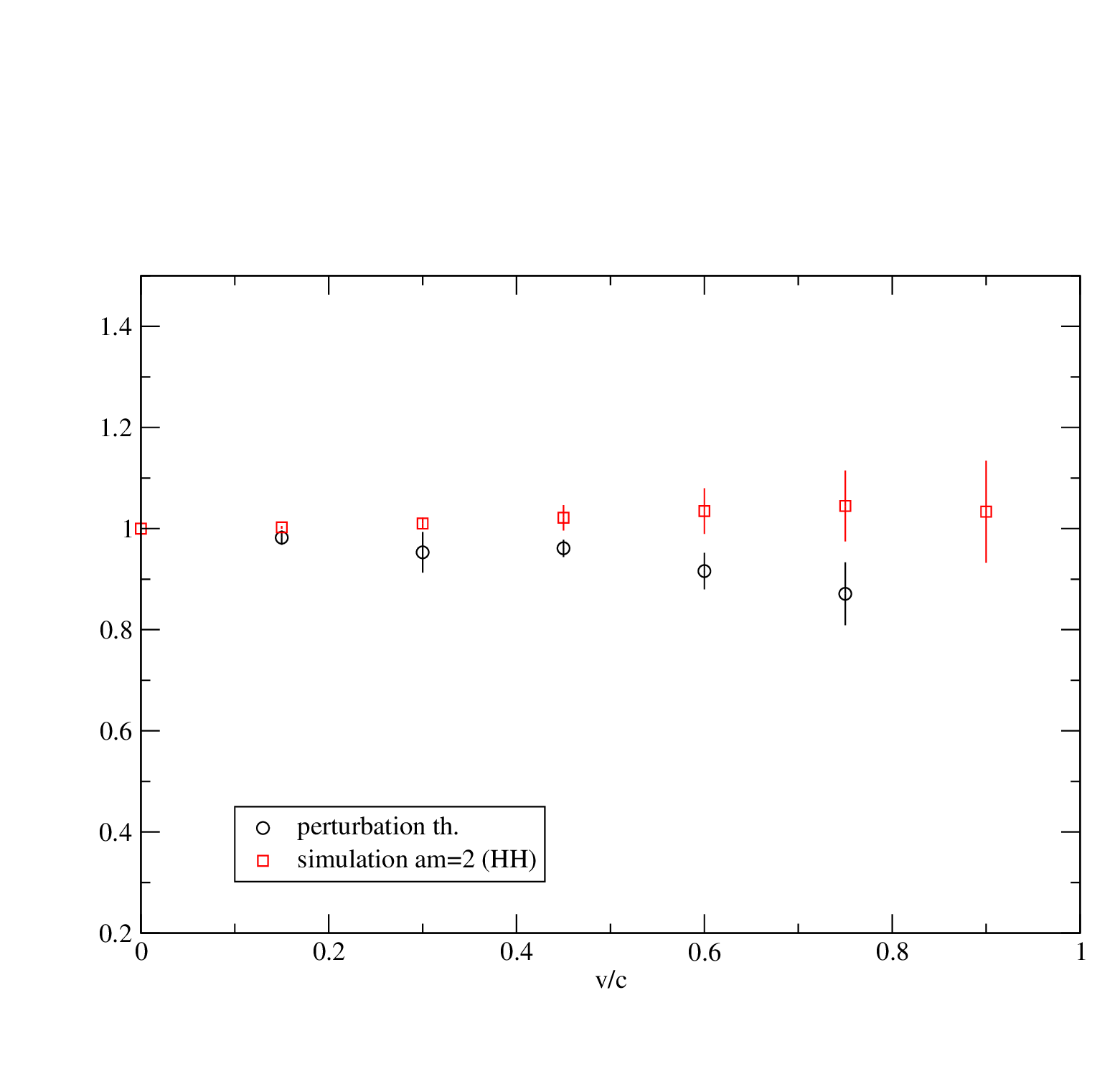}
      \caption{Comparison between the simulated and
        perturbative velocity renormalisation in mNRQCD.} 
      \label{zp}
    \end{center}    
    
  \end{minipage}
  \hspace{3mm}
  \begin{minipage}[r]{0.48\linewidth}
    \vspace{-1.1cm}
    \begin{center}
      \includegraphics[width=0.95\linewidth]{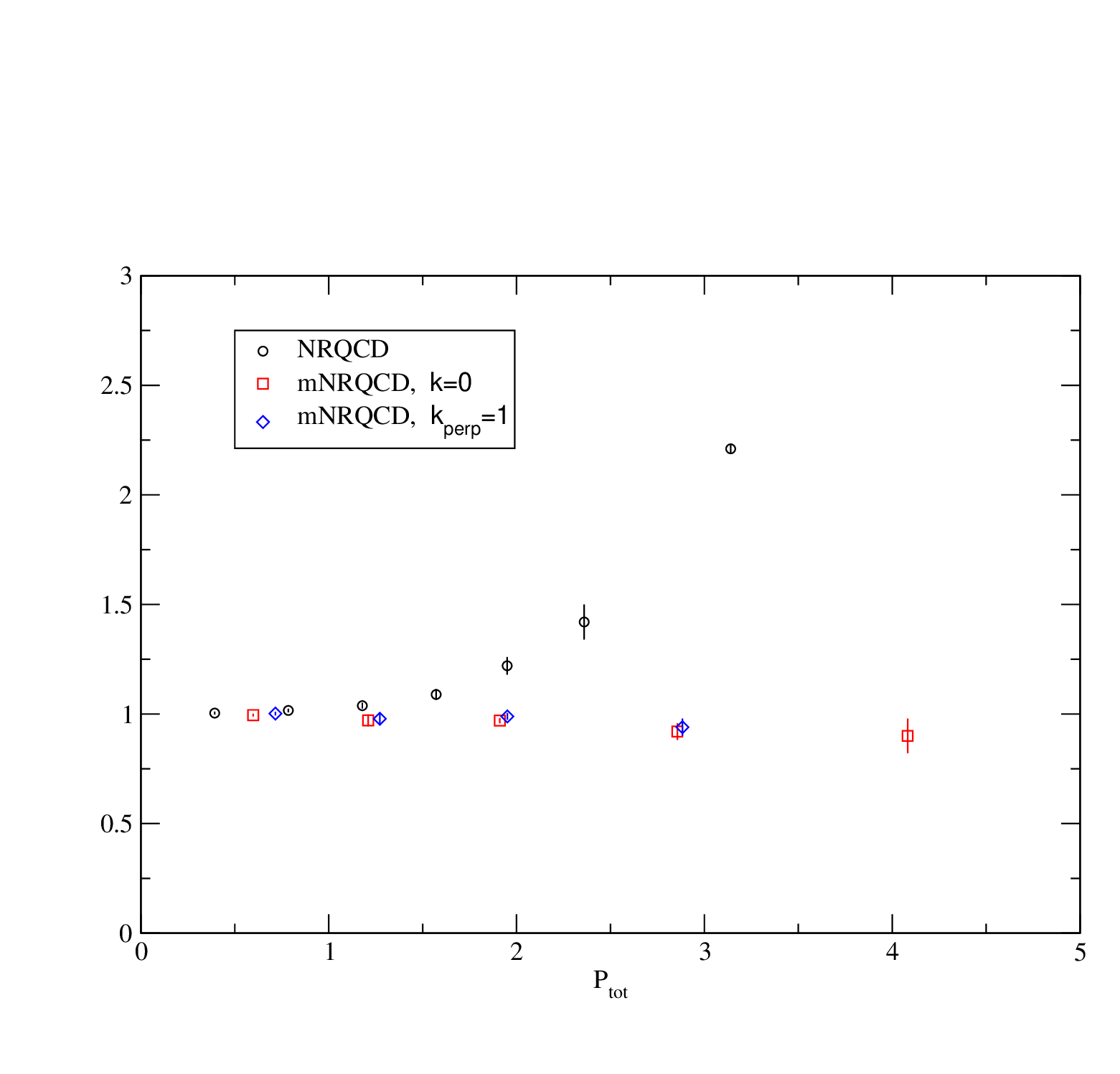}
      \caption{Comparison between the $\eta_B$ decay constant from the
        $A_0$ channel in NRQCD and mNRQCD. The mNRQCD data is plotted
        for two different values of the the residual momentum.} 
      \label{mnrqcd_fQQ_A0}
    \end{center}    
    
  \end{minipage}
\end{figure}

\begin{figure}
\vspace{-1cm}
  \begin{minipage}[l]{0.48\linewidth}
    \begin{center}
      \includegraphics[width=0.95\linewidth]{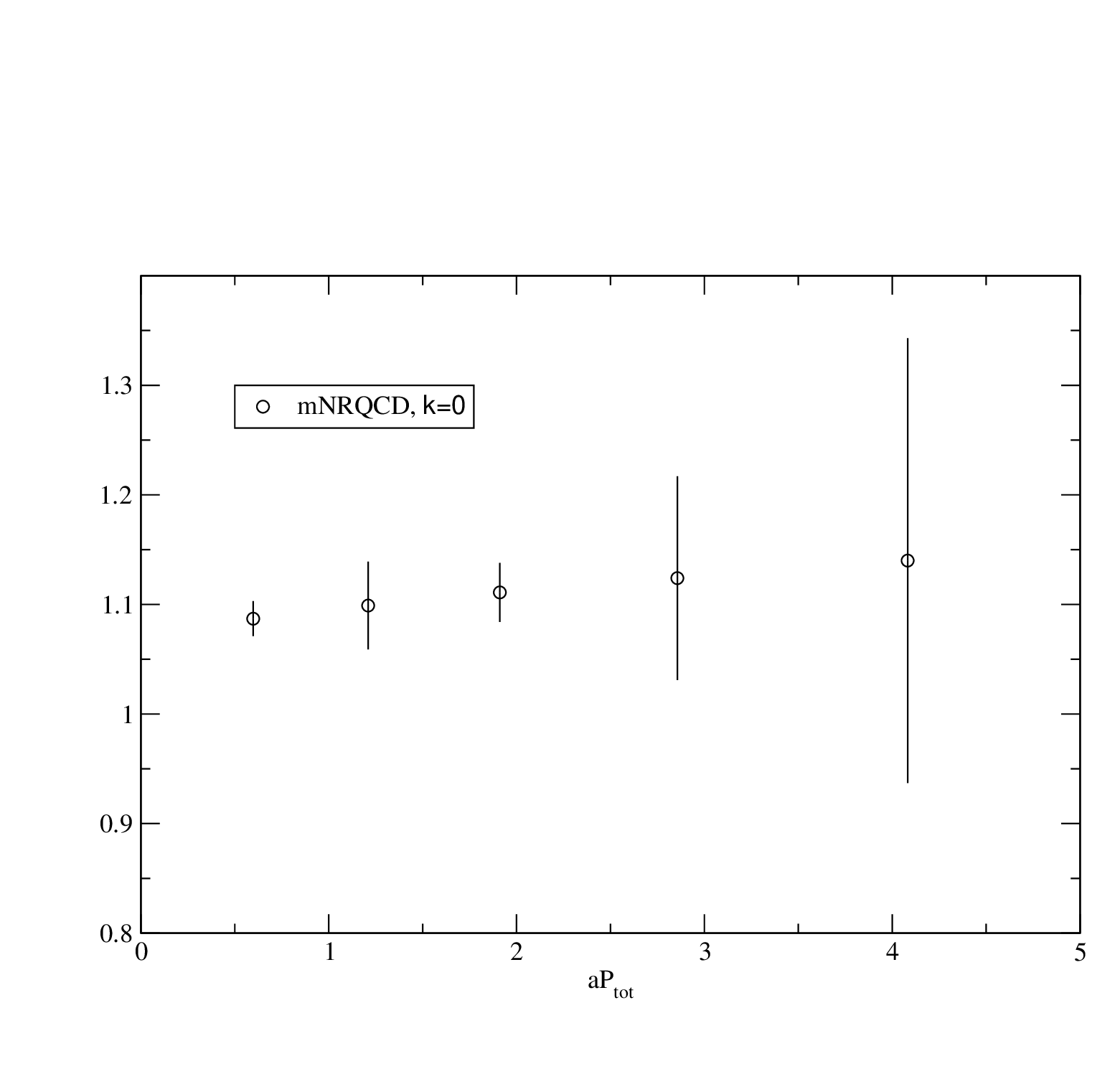}
      \caption{The ratio of the $\eta_B$ decay constant from the $A_k$ 
        channel and the $A_0$ channel. Note there is no leading
        order term in the NRQCD case, so the leading order answer is
        zero.}    
      \label{mnrqcd_fQq_Ak}
    \end{center}    
    
  \end{minipage}
  \hspace{3mm}
  \begin{minipage}[r]{0.48\linewidth}
    \vspace{-1cm}
    \begin{center}
      \includegraphics[width=0.95\linewidth]{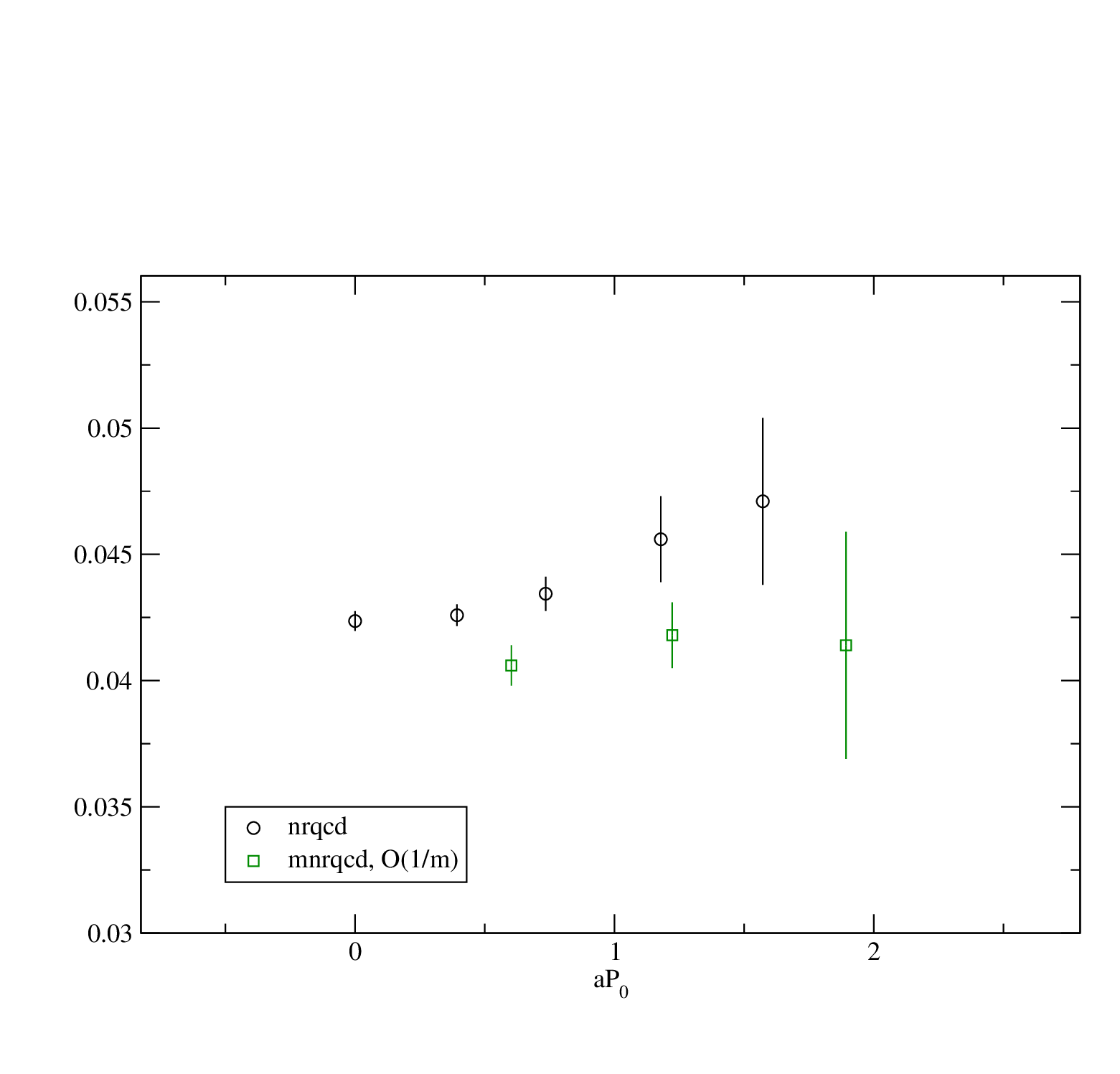}
      \caption{Comparison of the HH hyperfine splitting in
        NRQCD and $O(1/m)$ improved mNRQCD.}  
      \label{hyp}
    \end{center}    
    
  \end{minipage}
\end{figure}

There are a variety of quantities that can be computed (for both
heavy-heavy (HH) and heavy-light (HL) mesons) at different
values of the bare velocity in order to check that mNRQCD is working
correctly and that the renormalisations, such as $Z_P$, are under
control. Such quantities include the kinetic mass, the binding energy,
decay constants and energy splittings. The kinetic mass,
$M_{\mathrm{kin}}$ is extracted from the dispersion relation 
\begin{eqnarray*}
  E_v(k)+ C= \sqrt{(Z_P{\bf P}_0 + {\bf k})^2 + M_{kin}^2}
\end{eqnarray*} 
where ${\bf P}_0=\gamma m {\bf v}$. Figure \ref{mkin2} shows that the
kinetic mass (for the HH meson) is fairly constant as a function of
the bare velocity, an important feature as it means that the bare
quark mass will not have to be retuned to give the correct quark mass as
the velocity increases. In figure \ref{Esim}, we plot the ground state
energy, $E_{\mathrm{sim}}$ and the binding energy
($=E_{\mathrm{sim}}-E_0$), where $E_0$ is calculated from perturbation
theory. The velocity dependence has largly been removed in the
subtracted case.  In figure \ref{zp}, we compare the non-perturbative
and perturbative values for the momentum renormalisation $Z_P$. The
non-perturbative result uses the dispersion relation to compute $Z_P$
from $E_{\bf v}({\bf k})$. Figure \ref{mnrqcd_fQQ_A0} is a  plot of
the $\eta_B$ decay constant from the $A_0$ channel from both NRQCD and
mNRQCD. The plot is a ratio of the decay constant at varying total
momentum ($P_{\mathrm{tot}}$) over zero $P_{\mathrm{tot}}$. In the
NRQCD case, the discretisation errors become dominant at very low
$P_{\mathrm{tot}}$, whereas the mNRQCD ratio is constant. Figure
\ref{mnrqcd_fQq_Ak} shows the same ratio but the numerator  comes from
the $A_k$ channel. This is slightly larger than one, but should come
down to one as relativistic corrections are included. Finally, in
figure \ref{hyp} we present a preliminary plot for the hyperfine
splitting in the HH case from the NRQCD and $O(1/m)$ improved mNRQCD
action. 
 

\vspace{-.05cm}
\section{Summary}

We have presented a number of quantities computed from mNRQCD that
demonstrate its applicability to high recoil processes in which the
heavy quark carries large external momentum. Future  work will focus
on computing the $B\rightarrow \pi$ form factor. Due to the
complicated nature of a 2-loop perturbative calculation, the intention
is to employ numerical techniques to obtain the associated
renormalisation coefficients.

\bibliographystyle{JHEP}
\bibliography{proc}

\end{document}